\documentclass[11pt,twoside]{article}
\usepackage{asp2006}
\usepackage{psfig}
\usepackage{epsf}
\usepackage{graphics}
\usepackage{lscape}
\def\edcomment#1{\iffalse\marginpar{\raggedright\sl#1\/}\else\relax\fi}
\reversemarginpar

\begin{document}
\title{Virtual Observatory: Science capabilities and scientific results}
\author{Evanthia Hatziminaoglou}
\affil{European Southern Observatory, Karl-Schwarzschild-Str. 2, 
Garching bei M\"{u}nchen, Germany}

\begin{abstract}
The virtual observatory (VO) is a collection of interoperable data 
archives, tools and applications that together form an environment in 
which original astronomical research can be carried out.
The VO is opening up new ways of exploiting the huge
amount of data provided by the ever-growing number of ground-based 
and space facilities, as well as by computer simulations. This 
presentation summarises a variety of scientific results spanning various 
fields of astronomy, obtained thanks to the VO, after highlighting 
its structure, infrastructure and various capabilities.
\end{abstract}

\vspace{-0.5cm}
\section{Introduction}
\label{sec:intro}

A Virtual Observatory (VO) is a collection of inter-operating data 
archives and software tools which utilize the Internet to form an 
environment in which original research can be conducted.
The VO is an international astronomical 
community-based initiative, whose main goal is to allow transparent 
and distributed access to data available worldwide.
This is achieved by developing and applying common standards and by 
ensuring the interoperability
between the various data collections, tools and services.  This
allows scientists to discover, access, analyze, and combine 
observational and model data from heterogeneous data collections 
in a coherent and user-friendly manner.

\section{The VO concept in modern Astronomy}
\label{sec:concept}

The Virtual Observatory is a concept that arose from the need
of modern-day astronomy to exploit the wealth of
multi-instrument multi-wavelength datasets already available or to
come very soon, in a quest to answer primordial questions such as
that of the origin of the Universe, the formation and evolution of galaxies,
the formation of stars and planets.

Even though single observations provide {\it instantan{\'e}s} of smaller or larger 
parts of the Universe at a given wavelength regime and/or at a corresponding 
look-back time, a more in depth understanding of the 
Universe, its history and future requires sampling
the whole electromagnetic spectrum and studying the different epochs. 
And this need generates the problem of accessing increasingly large amounts of
data in combination of the huge volumes accumulated over the years 
and their subsequent exploitation. 

\begin{figure}[ht]
\centerline{
\psfig{file=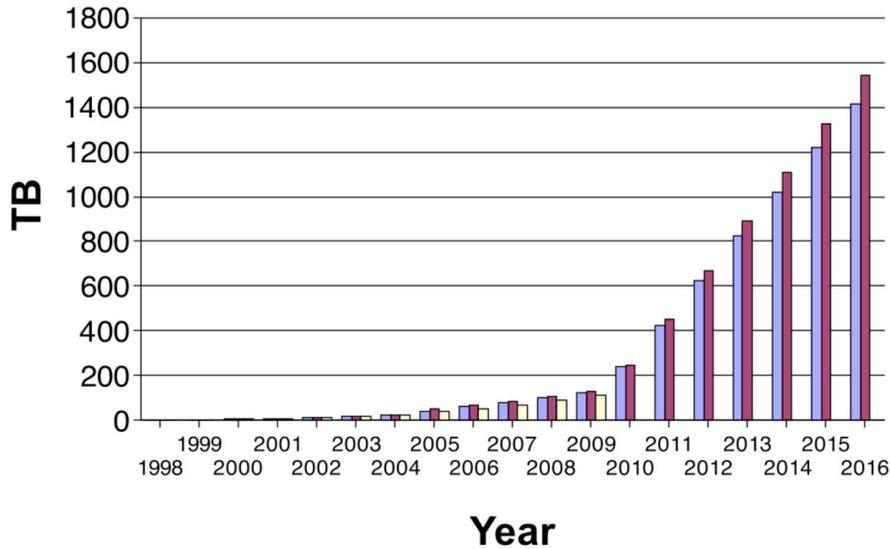,width=12.5cm}}
\caption{Past, present and projected future data volume in the ESO archive 
as a function of time (courtesy N. Fourniol; see also Eglitis \& Suchar 2009).}
\label{fig:esoarch}
\end{figure}

Modern observational astronomy is dominated by large multi-wavelength 
surveys producing TBs of data every night.
As an example of exploding data rates, Fig. \ref{fig:esoarch} shows the past and 
projected data volume in the ESO archive (in TBs) as a function of time.
These new data as well as the older data
are stored in {\it different} archives all over the world,
in {\it different} ways and formats, can be queried from {\it different} 
archive interfaces using {\it different} access methods and can be 
analysed with {\it different} tools and techniques.

In this framework, the VO is simply an organised effort to provide the 
science users a centralised and uniform access to all these data without 
having them worrying about different formats and access mechanisms, while 
simultaneously providing all the necessary information, in the form of 
metadata, for a scientific usage of these data collections.

The VO, however, is {\bf not} a centralised database of all the
astronomical data. It does not store the data - it merely provides
access to them. And it does {\bf not} play the role of data police.
The quality of the stored data is a responsibility of the data providers
and it is up to them to make sure that their quality and accompanying 
metadata are sufficient to ensure high-quality scientific research
based on their usage. Last but not least, the VO is not a monolithic
software system and there is no such thing as a ``killer application''.
There is not (and there will never be) a unique tool dealing with all the
aspects of imaging, spectral and tabulated data discovery, retrieval
and subsequent analysis. Instead, it provides a suite of multi-purpose 
tools and services and it is up to the users to select those that suit 
them best, based on their needs.

In its current status, the VO is in a transitional phase, going from 
development to operations. The amount of reduced data offered via VO services 
increases by the day, the tools, applications and services become more and more
mature and VO-enabled science is now taking place in parallel with
development, indicating the most urgent needs in term of data
availability, functionalities and standards. 

\section{The various VO entities}
\label{sec:vos}

\subsection{The International Virtual Observatory Alliance (IVOA)}
\label{sec:ivoa}

In order to insure the world-wide interoperability of applications and the
development and adoption of common standards, the International
Virtual Observatory Alliance (IVOA) was formed in June 2002.
The IVOA now comprises 17 VO projects from Armenia, Australia, Brazil, 
Canada, China, the European VO (EURO-VO), France, Germany, Hungary, 
India, Italy, Japan, Russia, South Korea, Spain, the United Kingdom, and the United States. 

The IVOA is structured in Working and Interest Groups, each one responsible of a given
area where key interoperability standards have to be defined and agreed upon.
An IVOA standard could be described as a set of guidelines and rules for data
providers or software developers who want their products (data or tools) to be
VO compliant and interoperable. It describes the information necessary for a
service or dataset to be VO compliant, both in terms of mandatory metadata
and implementation. IVOA standards are only relevant to data providers and
software developers and are meant to be completely transparent to the scientific
user.  

\subsection{The European Virtual Observatory (EURO-VO)}
\label{sec:eurovo}

The European Virtual Observatory (EURO-VO) project aims at deploying 
an operational VO in Europe. The EURO-VO is 
active in the areas of IVOA standards definition,
tools development, as well as dissemination for scientific usage.
Its internal structure consists of three bodies, namely the Data Centre Alliance,
the Technology Centre and the Facility Centre.

\begin{itemize}
\item the Data Centre Alliance, an alliance of 
European data centres, that provides
the physical storage and publishes data, 
metadata and services to the EURO-VO using VO technologies;
\item the Technology Centre is a distributed organisation 
that coordinates a set of technological projects related to standards,
systems and tools;
\item the Facility Centre, is the interface between
the EURO-VO and the astronomical community. Its immediate aims are to
introduce the VO capabilities to the European astronomers by means of
workshops, Schools and tutorials and to support research carried out using
the VO infrastructure.
\end{itemize}

\subsection{Other national VO initiatives}

Around the world, various national VO initiative are carrying out
activities in support of data publishing and scientific
usage of the VO. The US-VO\footnote{http://www.us-vo.org/}, for instance, offers 
a centralised portal where users can look for data in the VO as well
as exploit them with VO tools. The German VO (GAVO)\footnote{http://www.g-vo.org/}
offers, among other things, access to theoretical data (e.g. Millenium
simulation) through VO protocols.

\section{VO tools and services}
\label{sec:tools}

There is already a large variety of available VO-compatible 
applications for their immediate use to do science. Their level of 
maturity depends on a high degree on the level of maturity of the 
corresponding IVOA protocols and standards. As a consequence of the 
flexibility of the standards, as well as for historical reasons (i.e. 
many VO tools have evolved from existing pre-VO era applications) 
several of the applications might overlap in functionality.

Broadly speaking, VO tools can be classified in two main categories,
namely data discovery and data analysis, with most tools, however, suitable
for both categories.
A major advance towards interoperability, was the development
of SAMP, the Simple Application Messaging Protocol, that allows
applications to communicate with each other, building thus a complete
VO environment put together by various smaller pieces.
For an up-to-date list of VO tools, applications and services listed
either alphabetically or by functionality, see the section 
{\it Software} in the EURO-VO pages\footnote{http://www.euro-vo.org/software.html}.

\subsection{Data discovery}
\label{sec:discovery}

There is a variety of VO tools and services allowing for imaging,
spectra and catalogue discovery, such as Aladin\footnote{http://aladin.u-strasbg.fr/aladin.gml} 
developed by the CDS in France, 
Datascope\footnote{http://heasarc.gsfc.nasa.gov/cgi-bin/vo/datascope/init.pl}
provided by the US-VO,
or VODesktop\footnote{http://www.astrogrid.org/wiki/Home/AboutAGDesktop}
developed by AstroGrid, 
to name a few. They allow the user to look for all the
reduced data available in all the VO-compliant archives and data 
repositories (or {\it registries}) for a source,
browse through them, select those of interest and visualise in various forms. 
Most of the tools, in their current state and in interactive (i.e. non scripting) mode
allow either for single-object multi-resource query or for a
multi-object single-resource query.

\begin{figure*}[ht]
\centerline{
\psfig{file=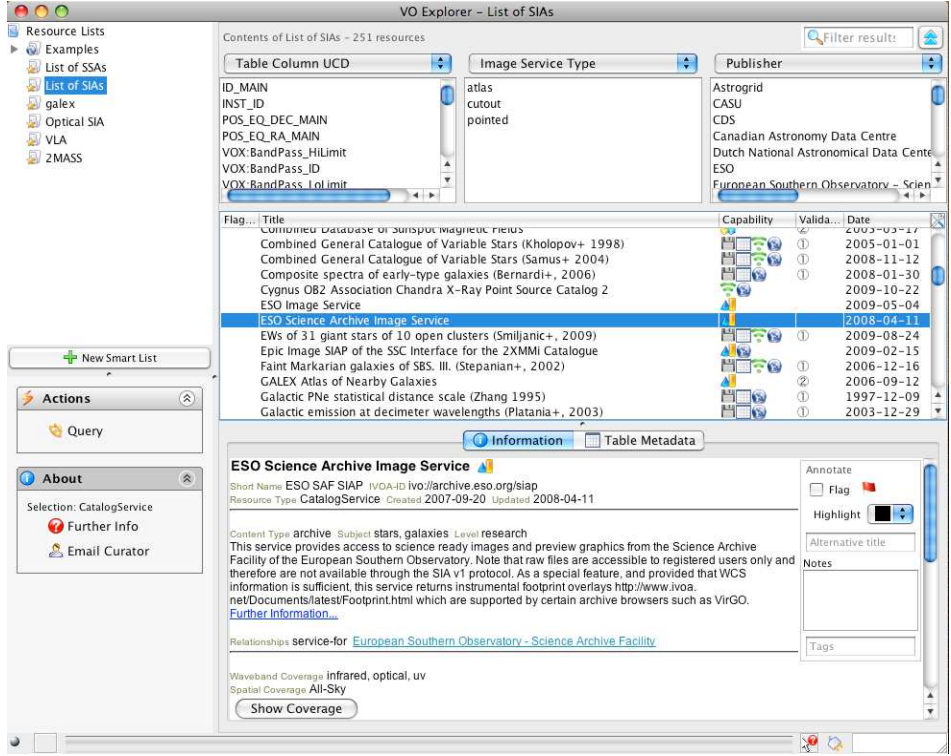,width=12.5cm}}
\caption{VODesktop main interface. For details on the various sections
see text.}
\label{fig:vodesktop}
\end{figure*}

Fig. \ref{fig:vodesktop} shows the VODesktop interface. In the top left the
list of resources grouped by user-defined criteria are shown. Each list
consists of a variety of resources, shown in the right part of the window.
For a selected resource, a detailed description is available at the bottom.
The services provided by each resource are listed at the right of the 
resource name (e.g. cone search is marked by a green cone, a globe indicates
a link to the web interface etc). The actions allowed for each selected resource are
listed at the bottom left of the window. Filtering the resources is
possible in a variety of ways, e.g. by service type, data Publisher,
column and many more.

Already existing tools are being adapted to exploit the VO world. An interesting
case is VirGO\footnote{http://archive.eso.org/cms/virgo/} 
\citep{hatzimi09a}, the new ESO Archive Browser. Since its latest release
(September 2009) VirGO comes with a collection of pre-configured VO data resources
(from ESO and external data centers like HST, CADC, ESA etc). A new configuration 
panel gives the possibility to add new ones in a quick and straight-forward way. And like every
other VO tool, VirGO now ``speaks'' SAMP, allowing for communication with other
VO applications.

\subsection{Data analysis}
\label{sec:analysis}

\subsubsection{Imaging data analysis}
\label{sec:images}
\begin{figure*}[ht]
\centerline{
\psfig{file=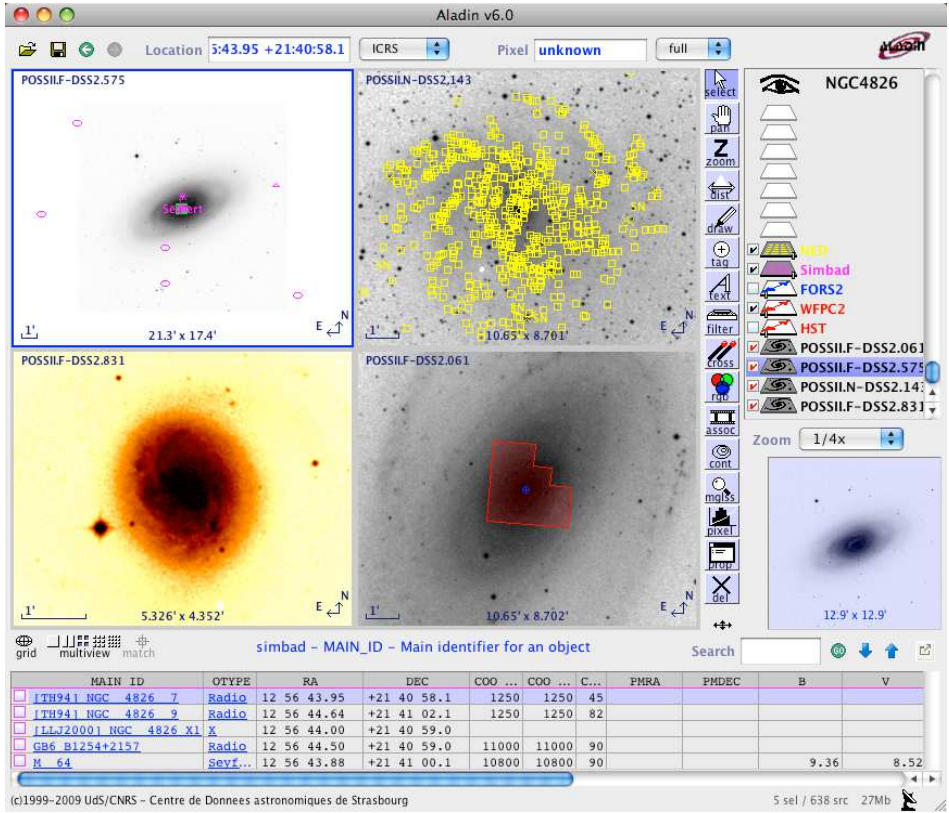,width=12.5cm}}
\caption{Aladin main window, showing in a multi-view mode from right to left
and top to bottom the following: NGC4826 with all SImbad objects in the
field superposed; NGC6946 with all NED objects in the field; NGC1068
with a modified colour map; and M81 with the WFPC2 footprint on it.}
\label{fig:aladin}
\end{figure*}

The most advanced imaging analysis VO tool to date is Aladin.
It is an interactive sky atlas that allows the
user to search, browse through and visualise images retrieved 
from VO resources, perform astrometric and photometric calibration
on them, superpose catalogues and instruments fields
of view, create colour-composite images and many more. It also includes a
scripting capability for multi-object (by target name or coordinates)
multi-resource search. Fig. \ref{fig:aladin} shows an Aladin
main window, showing in a multi-view mode from right to left
and top to bottom the following: NGC4826 Simbad objects in the
field superposed; NGC6946 with all NED objects in the field; NGC1068
with a modified colour map; and M81 with the WFPC2 footprint on it.

\subsubsection{Spectral analysis}
\label{sec:spectra}

Today the VO has three spectral analysis tools to offer, all three of them
very similar in functionalities, namely (in alphabetical order)
Specview\footnote{http://www.stsci.edu/resources/software\_hardware/specview/}
provided by the US-VO,
SPLAT-VO\footnote{http://star-www.dur.ac.uk/$\sim$pdraper/splat/splat-vo/}, provided
by AstroGrid, and 
VOSpec\footnote{http://www.sciops.esa.int/index.php?project=ESAVO\&page=vospec},
developed by ESA-VO.

\begin{figure*}[ht]
\centerline{
\psfig{file=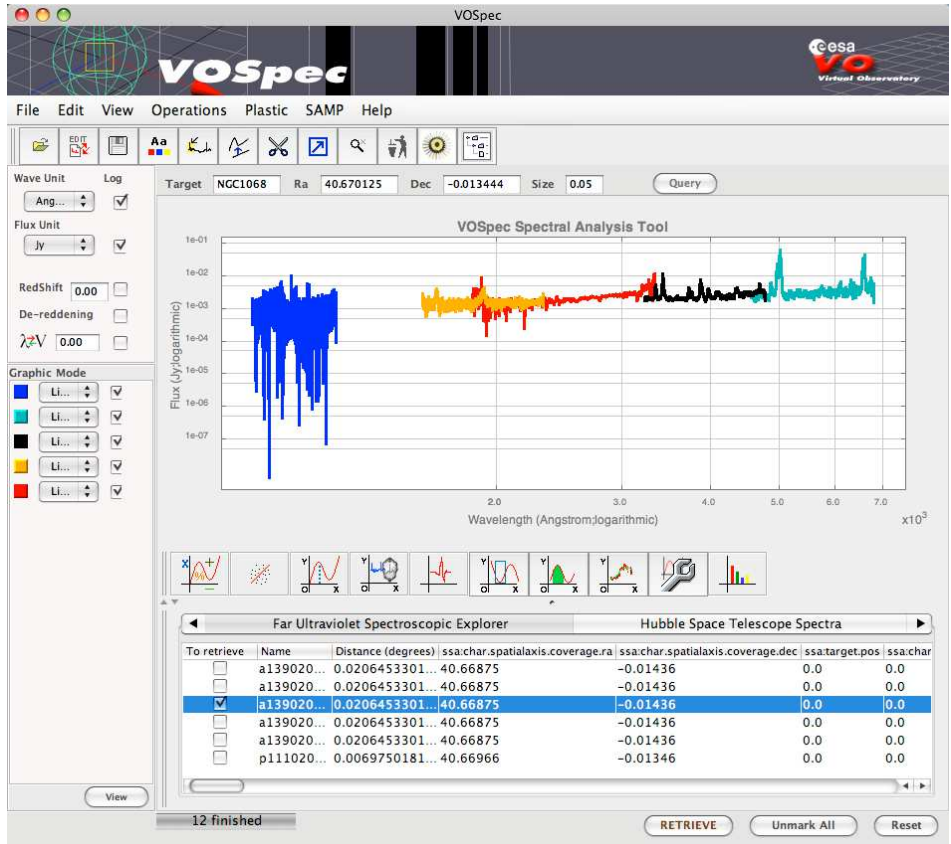,width=12.5cm}}
\caption{VOSpec screenshot showing the SED of NGC1068 built using FUSE, HST and ISO spectra
queried directly from relevant VO resources.} 
\label{fig:vospec}
\end{figure*}

All three tools allow the user to directly query all the available registries 
that contain ground-based and space spectral data and/or load local files. 
Other than the display capabilities, the tools provide various high level
analysis modules such as line measurement and fitting, line identification etc.
Fig. \ref{fig:vospec} shows a spectral energy distribution (SED) for NGC1068
built using FUSE, HST and ISO spectra, constructed with VOSpec.

\subsubsection{Handling of tabular data}
\label{sec:tables}

\begin{figure*}[ht]
\centerline{
\psfig{file=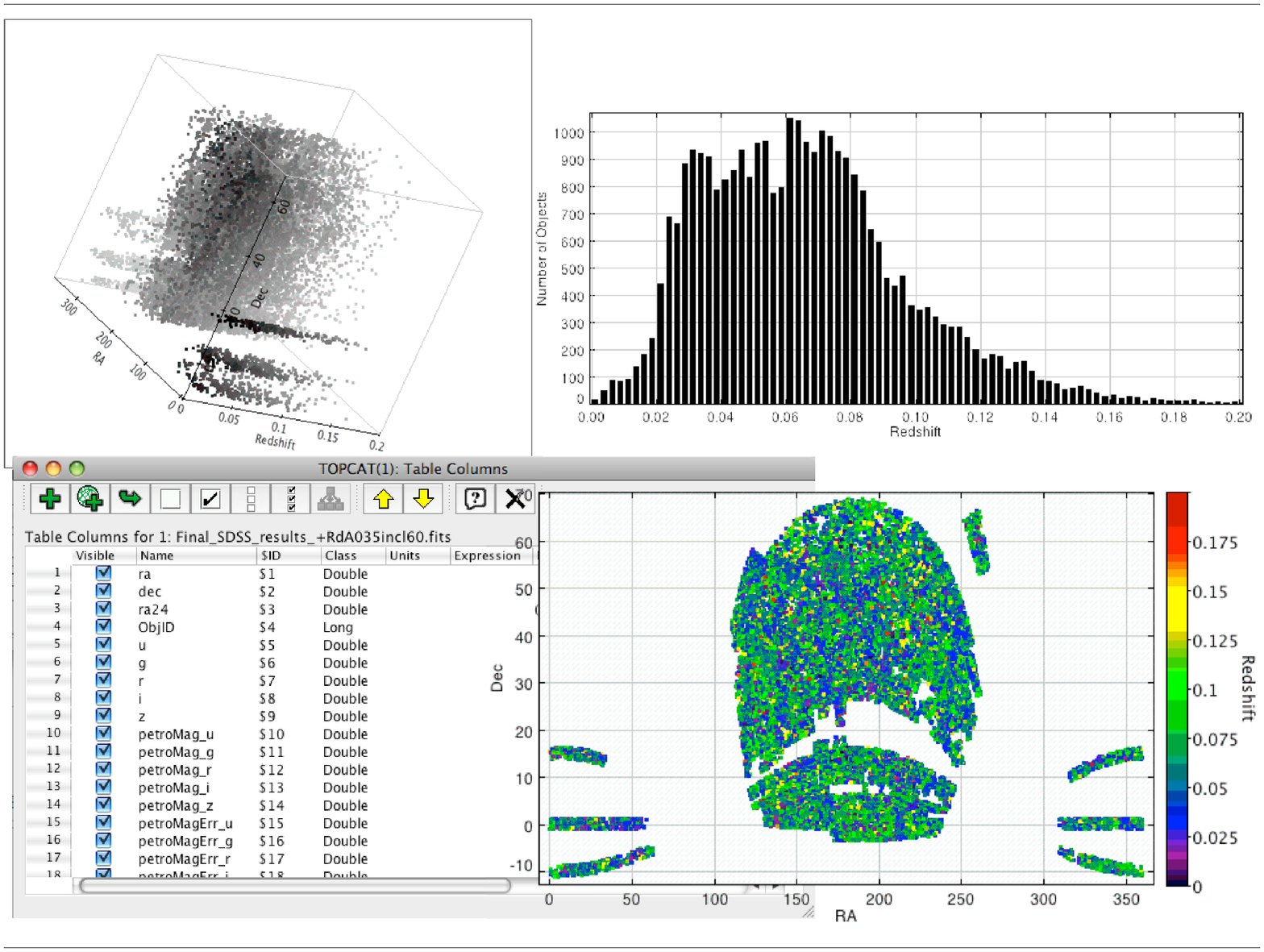,width=12.5cm,angle=-90}}
\caption{From left to right and top to bottom, a 3D plot (with RA, Dec and 
redshift as axes; a redshift histogram; the table column editor and a scatter
plot (RA, Dec) using a third column (redshift) for colour coding. A subsample
of the SDSS DR6 galaxy sample has been used for the creation of the plots.}
\label{fig:topcat}
\end{figure*}

A nice example of a tabular data handling tool is
TOPCAT\footnote{http://www.star.bris.ac.uk/$\sim$mbt/topcat/},
distributed as part of the AstroGrid tool suite.
TOPCAT offers a wide variety of data visualisation functionalities,
such as histogrammes, scatter plots, 3D and density plots;
table and column metadata visualisation and editing (e.g. operation
among columns, coordinate transformations); table format conversions
(supported formats include ascii, fits, VOTable, tsv and csv [tap- and 
coma-separated values tables, respectively], latex and more); cross-correlation
capabilities of up to four tables simultaneously using various
cross-correlation algorithms and many more. It also uses VO standards
in order to query the VO-compliant resources and therefore allows the user to
combine data residing on a local disk with data available through
the VO. Thanks to the implementation of SAMP, TOPCAT can also
communicate with all other VO tools and applications.
Fig. \ref{fig:topcat} shows a collection of screenshots of the above-mentioned
capabilities. 

For tables larger than a few tens of thousands of rows, 
STILTS\footnote{http://www.star.bris.ac.uk/$\sim$mbt/stilts/},
the STIL Tool Set,
a command-line tool is recommended. STILTS performs pretty
much the same operations TOPCAT does but in a non-interactive, scriptable
way. It is also recommended in case of operations having to be repeated
several times.

VOPlot\footnote{http://vo.iucaa.ernet.in/$\sim$voi/voplot.htm} is an 
alternative to TOPCAT, developed by VO-India, capable of operating
on tables in VOTable format. Aladin has also some table querying
and handling capabilities. It allows to query the VO registry for all
available tabular datasets based on coordinates or target names,
and allows for cross-correlations. Cross-correlations of up to three
tables simultaneously and up to 5000 objects at a time can also be
performed using Open SkyQuery\footnote{http://www.openskyquery.net/Sky/skysite/OSQform/default.aspx}
developed by the US-VO.

\section{Recent scientific results}
\label{sec:science}

The number of 
VO-enabled publications is now rising and the research carried out is covering
all aspects on modern astronomy, from solar and stellar physics to extragalactic
astronomy and cosmology.
The main goal of the VO has been to allow for a seamless data access
across archives; and even though VO tools and infrastructure are mostly
used as a platform for data mining, other types of usage such as using VO tools
for combined data visualisation, has been seen in
the recent literature. 

Ever since the first, fully VO-enabled discovery of 31 type 2 QSO candidates in the
GOODS fields \citep{padovani04}, many new discoveries were made with the help of 
VO applications and services. An exploitation of the IPHAS catalogue \citep{gonzalez08} in
combination with 2MASS in an area
of 1600 square degrees recently revealed 23 accreting young, low-mass stars and brown 
dwarfs far from known star-forming regions \citep{valdivielso09}, 
suggesting the existence of hundreds of such objects
in the area covered by IPHAS. Similarly, a high proper
motion L sub-dwarf was recently identified in the SDSS catalogue combining SDSS
and 2MASS spectrophotometric information \citep{sivarani09}.
Taking advantage of the wide field
of view of HRI/ROSAT and the availability of public data through VO tools, 
\cite{caballero09} investigated the daily X-ray variability of 23 stars in the young
$\sigma$~Orionis cluster. They detected unusual flares with durations of about six days,
duration considerably larger than the $\sim$24 hours usually observed in active stars
residing in very young star forming regions. This last work is actually part of an
ongoing investigation of the $\sigma$ Orionis region that also revealed large numbers
of previously unknown brown dwarf candidates by pasting together various large
photometric catalogues (\citealt{caballero08a}; \citealt{caballero08b}).

A major contribution of the VO to astronomy is the possibility it offers for
retrieving, exploiting and visualising large datasets. These capabilities have been
extensively used lately to produce new results. In \cite{fathi09} for instance, 
a sample of $\sim$55000 Sa-Sd SDSS galaxies was selected, cross-correlated with 
HyperLeda\footnote{http://leda.univ-lyon1.fr/}
in order to extract the morphological information and then {\it u}, {\it g}, {\it r}, 
{\it i}, {\it z}, {\it J}, {\it H} and {\it K} band images for each one of them were
retrieved from SDSS and 2MASS using various VO services, with the aim of studying the
scale length of disk galaxies as a function of redshift and morphology.
Large catalogues of legacy value constructed with the help of the VO infrastructure
have also been published recently. One example is the Large Quasar Astrometric
Catalogue (LQAC; \citealt{souchay09}) that contains 113666 quasars with photometry
in {\it u}, {\it b}, {\it v}, {\it g}, {\it r}, {\it i}, {\it z}, {\it J}, {\it K} 
as well as redshift and radio fluxes at 1.4, 2.3, 5.0, 8.4 and 24 
GHz, when available. This catalogue is the results of the combination of the 12
largest quasar catalogues available in the literature.

The volume of datasets available through
VO protocols is increasing exponentially, thanks to many large projects
and surveys providing their data through VO-compliant interfaces but also thanks
to the smaller projects that make the extra effort of adding a ``VO-layer"
to their data and publishing them using VO protocols. 
Recently, the Hubble Legacy Archive (HLA) made available through VO interfaces
all HLA NICMOS G141 grism spectra \citep{freudling08}, enabling thus
flexible querying of the 2470 available spectra.
An example of feedback from the community to the VO
was the effort off putting together a catalogue of quasar candidates selected from
their photometry extracted from the SDSS DR7, presented in \cite{dabrusco09}.
The two methods used for the quasar candidates selection, namely the probabilistic 
principal surfaces and the negative entropy clustering, were subsequently implemented 
into the AstroGrid VO platform.

Smaller and larger projects are also using the VO tools for their visualisation
capabilities. Such an example is Exo-Dat, the information system created to support
CoRoT \citep{deleuil09}. 
Exo-Dat has implemented various IVOA protocols and output formats and
allow for a direct interfacing with VO tools such as Aladin and TOPCAT,
avoiding thus having to develop its own visualisation tools that would overlap
in functionality with that of the existing VO applications.

\section{Conclusions}
\label{conclude}

The scientific usage of the VO tools and infrastructure is now taking up,
and this is reflected by the increasing number of refereed and other publications
making use of the VO capabilities. 
The EURO-VO and other national VO initiative are here to help interested groups
to carry out VO-enabled research and provide support for projects that want to
publish their data in the VO. Dedicated initiatives such a workshops and
Schools are organised world-wide in order to make astronomers aware of the
grand potential of the VO and the scientific possibilities it offers.

\end{document}